\documentstyle[11pt,aaspp4]{article}
\begin{document}
\title{Towards a Model for the Progenitors of Gamma-Ray Bursts\\
~\\
\it{In memory of Jan van Paradijs}\\
}
\author{Mario Livio\altaffilmark{1} and Eli Waxman\altaffilmark{2}}
\altaffiltext{1}{Space Telescope Science Institute, 3700 San Martin Drive,
Baltimore, MD 21218, USA}
\altaffiltext{2}{The Weizmann Institute of Science, Rehovot 76100, Israel}

\begin{abstract}
We consider models for gamma-ray bursts in which a collimated jet expands 
either into a homogeneous medium or into a stellar wind environment, and 
calculate the expected afterglow temporal behavior. We show that (i)~following 
a break and a faster decay, afterglows should exhibit a flattening,
which may be detectable in both the radio and optical bands; (ii)~Only 
observations at times much shorter than a day can clearly distinguish between a 
fireball interacting with a homogeneous medium and one interacting with a 
stellar wind. 

Using our results we demonstrate that constraints can be placed 
on progenitor models. In particular, existing data imply that while
some long duration bursts may be produced by collapses of massive stars,
it is almost certain that not all long duration bursts are produced by such
progenitors.
\end{abstract}

\section{Introduction}

During the past three years the understanding of gamma-ray bursts (GRBs) has
literally been revolutionized. The discovery of x-ray afterglows by BeppoSAX
(e.g.\ Costa et~al.\ 1997) followed by the discovery of optical transients
(e.g.\ van Paradijs et~al.\ 1997), led eventually to a full confirmation of the
cosmological nature of (at least a subclass of) GRBs.  The latter has been
achieved both by direct redshift measurements (e.g.\ Metzger et~al.\ 1997), and
by imaging of the host galaxies (e.g.\ Sahu et~al.\ 1997).

Since GRBs involve the generation of huge amounts of energy during very short
time intervals, most GRB models involve compact or collapsed objects (e.g.\
Paczynski 1986; Eichler et~al.\ 1989; Fryer, Woosley \& Hartmann 1999; and see
M\'esz\'aros 1999 for a review). In spite of impressive successes of expanding
relativistic ``fireball'' models (e.g.\ Paczynski \& Rhoads 1993; Katz 1994;
M\'esz\'aros \& Rees 1997; 
Vietri 1997; Waxman 1997a; Sari, Piran \& Narayan 1998), the
precise nature of GRB progenitors remains unknown.

In recent years it has become clear that, like in supernovae, GRB progenitors
may in fact span a rather heterogeneous class. At least two broad groups are
evident from plotting spectral hardness versus burst duration (e.g.\ Katz \&
Canel 1996; Kouveliotou et~al.\ 1996; Fishman 2000). One group consists of
relatively hard and
short (mean duration of $\sim0.2$~sec) bursts, while the other of softer and
longer (mean duration of $\sim20$~sec) bursts. All the afterglows and optical
transients discovered so far followed bursts belonging to the second group. 

On the basis of the burst durations and the estimated burst frequencies it is
generally speculated (e.g.\ Fryer et~al.\ 1999) that the short duration bursts
are the results of mergers, mostly of neutron star--neutron star 
(NS--NS) and black hole--neutron star (BH--NS) pairs, while the long duration
bursts are mostly the results of collapses of massive stars. The latter scenario
received some support from the tentative identification of GRB~980425 with the
Type~Ic supernova SN~1998bw (Galama et~al.\ 1998; Kulkarni et~al.\ 1998; Iwamoto
et~al.\ 1998), and from a tentative detection of a supernova underlying
GRB~980326 (Bloom et~al.\ 1999).

In the present work we attempt to take the identification of GRB progenitors one
step further, by examining in some detail the behavior of GRB afterglows. In
particular, we investigate expected (and observed) breaks in the power-law
decline of afterglows, and their potential relation to jets and to interaction
with a pre-outburst stellar wind environment. The general framework and
calculations are presented in \S2 and a discussion and conclusions follow.

\section{Jets and Winds}

An examination of the temporal decay of the afterglows of GRBs reveals the
following three trends:
\begin{enumerate}
\item[({\it i})] In a few afterglows (like GRB~970228, GRB~970508) the decay
broadly followed a single, unbroken power law, behaving like $t^{-1.14\pm0.05}$
and
$t^{-1.23\pm0.04}$ for the above two respectively, (e.g.\ Fruchter et~al.\ 1998;
Zharikov, Sokolov \& Baryshev 1998; although see Reichart 1999, Galama et~al.\
1999).
\item[({\it ii})] In some GRBs (like GRB~990123 and GRB~990510) the optical
afterglow decayed like one power-law initially, and then began to decline
faster. For
example, GRB~990123 behaved like $t^{-1.1\pm0.03}$ from 3.5~hours after the
burst till about two days after the burst, when it started a steeper decline
(Kulkarni et~al.\ 1999). Similarly, GRB~990510 behaved like 
$t^{-0.76\pm0.01}$ at early times ($t\ll1$~day) and $t^{-2.40\pm0.02}$ at late
times (Stanek et~al.\ 1999).
\item[({\it iii})] In two GRBs (GRB~980519 and GRB~980326) the afterglow was
observed to fade very rapidly, like $t^{-2.05\pm0.04}$ and $t^{-2.1\pm0.13}$
respectively (e.g.\ Halpern et~al.\  1999; Groot et~al.\ 1998).
\end{enumerate}

Two types of potential explanations have been suggested for the break in (or
very fast decline of) the light curve. In one explanation, the fireball is
initially
{\it a highly collimated jet}, with the break occurring when the Lorentz
factor $\Gamma$ becomes smaller than 1/$\theta$, where $\theta$ is the jet
opening angle (e.g.\ Rhoads 1999; Panaitescu \& M\'esz\'aros 1998; Sari, Piran
\& Halpern 1999; Dar 1998; Harrison et~al.\ 1999; Stanek et~al.\ 1999). In the 
second model, it has been suggested that the steep
decline (e.g.\ of GRB~980519) was caused by the interaction of {\it a
spherical burst} with a pre-burst Wolf-Rayet star wind (e.g.\ Chevalier \& Li
1999a,b; M\'esz\'aros, Rees \& Wijers 1998; Frail et~al.\ 1999).

Let us first examine these two suggestions and their potential relation to
progenitor models. The first point to note is that collimation and acceleration 
of jets (in the context of MHD extraction of energy, as opposed to neutrinos) 
is generally thought to occur by an accretion disk which is threaded by a
large-scale vertical magnetic field (e.g.\ Blandford 1993; Spruit 1996; Livio 
1999; although purely hydrodynamical disk models have also been considered, 
e.g.\ MacFadyen \& Woosley 1999). This model has received additional recent 
support from very high-resolution VLBI observations of the M87 jet (Junor, 
Biretta \& Livio 1999), which show the collimation process occurring on 
scales of 30--100 Schwarzschild radii from the putative central black hole, 
with the jet exhibiting limb-brightening in the radio. Good 
collimation requires a relatively large ratio,
$R_d/R_{CO}\gg1$, (where $R_d$ is the disk outer radius and $R_{CO}$ is the
radius of the central compact object). This condition is naturally satisfied in
the case of a collapsing massive star, but not in the coalescence of NS--NS or
BH--NS binaries (where $R_d/R_{CO}\sim1$). Hence, while highly collimated jets
may be expected from massive star collapses (in which case high collimation is
also needed for the jet to be able to escape the collapsing mantle), collimation
is probably at best moderate from NS--NS and BH--NS mergers.

Secondly, it is virtually {\it impossible} to avoid the existence of a
stellar wind environment in massive star GRB progenitor models. The wind 
mass-loss rates from Wolf-Rayet stars are approximately of order
$\dot{M}\sim6.3\times
10^{-6}(M_{WR}/10~\rm{M}_{\odot})^{2.5}$~M$_{\odot}$~yr$^{-1}$ 
(e.g.\ Hamann \& Koesterke 1998), and they do not depend very significantly 
on metallicity (e.g.\
Willis 1991; Leitherer 1991; in addition, most GRBs with afterglows so far are
at redshifts $z\la1$, so the change in the cosmic metallicity is not dramatic).

The above two points indicate immediately the following consequences: 
\begin{enumerate}
\item[({\it i})]Since {\it all} the observed afterglows belong to the
long-duration group of GRBs, {\it if} some afterglows do not show any clear 
signs of either highly collimated jets or interaction with a wind (e.g.\ in
terms of breaks, fast declines, or radio evolution), then the group of 
long-duration GRBs is almost certainly not {\it all} resulting from
collapses of massive stars.
\item[({\it ii})]Since GRBs resulting from collapses of massive stars lead
{\it both} to highly collimated jets (at least sometimes) and to a stellar
wind 
environment (always), it is important to examine the development of a jet 
interacting with a wind.
\end{enumerate}

Based on the above two points, we reexamine now the behavior expected from
collimated jets.

\subsection{Jet Transition to Sub-Relativistic Expansion}

We first consider jets expanding into a homogeneous medium. The discussion
is generalized in \S2.2 to jets expanding into a wind. 

As long as the jet Lorentz factor $\Gamma$ is larger than the inverse of the
jet's opening angle $\theta$, it behaves as if it were a conical section of 
a spherical fireball. At this stage, the Lorentz factor as a function of 
jet radius is given by the Blandford-McKee (1976) solution, $\Gamma=(17E_i/16\pi 
n m_p c^2)^{1/2}r^{-3/2}$, where $n$ is the number density ahead of the shock 
and $E_i$ is the energy the fireball would have carried if it were spherically
symmetric, i.e. the jet energy is 
$E=\theta^2E_i/2$ (for a double sided jet).
Once $\Gamma$ drops below
$\theta^{-1}$, the jet expands sideways, and its behavior deviates from the
spherically symmetric
case. We define $t_\theta$ as the time, as measured by a distant observer, at
which $\Gamma=1/\theta$. Using the Blandford-McKee solution, and the relation
$t=r/4\Gamma^2c$ between observer time and jet radius
(Waxman 1997b), we find 
\begin{equation}
\theta\approx\left({17\over 1024\pi}{E_i\over n m_p c^5}\right)^{-1/8}
       t_\theta^{3/8}=
       0.12\left({E_{i,53}\over n_0}\right)^{-1/8}t_{\theta,\rm day}^{3/8}~~,
\label{eq:theta}
\end{equation}
where $E_i=10^{53}E_{i,53}$~erg, $n=1n_0$~cm$^{-3}$. 

After a transition stage, in which the jet expands sideways, the flow approaches
spherical symmetry, and can again be described by a simple self-similar
solution. The transition takes place over a time $t_s\approx r_\theta/c$,
where  $r_\theta$ is the jet radius at time $t_\theta$. Thus,
\begin{equation}
t_s\approx r_\theta/c=270
\left({E_{i,53}\over n_0}\right)^{1/4}t_{\theta,\rm day}^{1/4}\ {\rm day}~~.
\label{eq:t_s}
\end{equation}
It is straightforward to show that after the transition
the flow becomes sub-relativistic (e.g.\ Waxman, Kulkarni \& Frail 1998). 
At time $t_\theta$ the energy associated with the mass enclosed within a 
sphere of radius equal to the jet radius $r_\theta$, $M_\theta\equiv4\pi 
n m_p r_\theta^3/3$, is similar to the jet total energy, 
$E/M_\theta c^2=6/17$. 
Thus, after
spherical symmetry is approached, the flow is described by the 
Sedov-von Neumann-Taylor solutions. 

The fireball radius during the sub-relativistic expansion is given by the
Sedov-von~Neumann-Taylor relation, $r=\xi(Et^2/n m_p)^{1/5}$ where
$\xi\approx1$ depends on the gas adiabatic index. We define 
the time $t_{\rm SNT}$ as the time at which $\dot r_{\rm SNT}/c=1$. For
$\xi=1$ we have
\begin{equation}
t_{\rm SNT}=67
\left({E_{i,53}\over n_0}\right)^{1/4}t_{\theta,\rm day}^{1/4}\ {\rm day}~~.
\label{eq:t_SNT}
\end{equation}

The transition from a jet to a Sedov-von~Newman-Taylor behavior should occur
during the period $t_{\rm{SNT}}\la t \la t_s$, with noticeable deviations
from the collimated jet behavior starting at $t\sim t_{\rm{SNT}}$.

After the transition, the power is dominated by synchrotron emission from a
sub-relativistic fireball (e.g.\ Frail, Waxman \& Kulkarni 1999). 
Assuming that a fraction $\xi_e$ ($\xi_B$) of the thermal
energy behind the shock is carried by electrons (magnetic field), and that
the electrons are accelerated to a power-law energy distribution,
$dn_e/d\gamma_e\propto\gamma_e^{-p}$ with $p=2$, the flux at frequencies above
\begin{equation}
\nu_*\approx1\left({1+z\over2}\right)^{-1}\left({\xi_e\over0.3}\right)^2
\left({\xi_B\over0.3}\right)^{1/2}n_0^{1/2}\,{\rm GHz}
\label{eq:nu_m}
\end{equation}
is given by (see e.g.\ Appendix of Frail, Waxman \& Kulkarni 1999)
\begin{equation}
f_\nu\approx1\left({1+z\over2}\right)^{-1/2}\left({\xi_e\over0.3}\right)
\left({\xi_B\over0.3}\right)^{3/4}n_0^{3/4}E_{51}d_{28}^{-2}
\nu_{\rm GHz}^{-1/2}\left({t\over t_{\rm SNT}}\right)^{-9/10}\,{\rm mJy}~~.
\label{eq:f_nu}
\end{equation}
Here, $d_L=(1+z)^{1/2}10^{28}d_{28}$~cm and $\nu_*$ is the synchrotron peak 
frequency at $t=t_{\rm SNT}$. Eq. (\ref{eq:f_nu}) is valid for frequencies where
emission is dominated by electrons with cooling time larger than the expansion
time. At higher frequencies, above
\begin{equation}
\nu_c\approx10^{13}{2\over1+z}
\left({\xi_B\over0.3}\right)^{-3/2}n_0^{-5/6}E_{51}^{-2/3}
\left({t\over t_{\rm SNT}}\right)^{-1/5}\,{\rm Hz}\quad,
\label{eq:nu_c}
\end{equation}
the spectrum steepens to $f_\nu\propto\nu^{-1}$.

Eqs. (\ref{eq:nu_m}--\ref{eq:nu_c}) imply 
that at $t\sim t_{\rm SNT}$, radio emission at 
the 1~mJy level is expected. 
While the optical flux at this stage is of the order of
$1\mu$Jy, and thus not easy to detect, it is not altogether undetectable.

\subsection{Jet-Wind Interaction}

Let us now consider a fireball jet expanding into a wind, where
the ambient medium density is $\rho=\dot M/4\pi v_{\rm w} r^2$ (where 
$v_{\rm w}$ is the wind speed). During the
stage in which $\Gamma>1/\theta$, we may obtain an approximate description
of the dynamics by using the $r$ dependent number density $n(r)=Ar^{-2}$ in
the Blandford-McKee relation for $\Gamma(r,E,n)$. Using this relation,
and $t=r/4\Gamma^2c$, we find that the density (ahead of the shock)
at observed time $t$ is
\begin{equation}
n\approx{4\pi\over17}cA^2(E_i m_p t)^{-1}=
0.4\left({\dot M_{-5}\over v_{{\rm w},3}}\right)^2E_{i,53}^{-1}
t^{-1}_{\rm day}\ {\rm cm}^{-3},
\label{eq:n}
\end{equation}
where $\dot M=10^{-5}\dot M_{-5}$~M$_\odot/{\rm yr}$ and 
$v_{\rm w}=10^3v_{{\rm w},3}$~km/s. The relation between the jet
opening angle and the time at which a deviation from spherical behavior
is observed, Eq.~(\ref{eq:theta}), may be generalized to the wind
case by replacing $n$ (in Eq.~(1)) with $n(t)$ given by Eq.~(\ref{eq:n}), 
to give
\begin{equation}
\theta\approx0.11\left({E_{i,53}\over\dot M_{-5}/v_{{\rm w},3}}\right)^{-1/4}
t^{1/4}_{\theta,{\rm day}}.
\label{eq:theta-jet}
\end{equation}
Similarly, for the wind case $t_{\rm SNT}$ is given by
\begin{equation}
t_{\rm SNT}=69
\left({ E_{i,53}\over\dot M_{-5}/v_{{\rm w},3}}\right)^{1/2}
t_{\theta,\rm day}^{1/2}\ {\rm day}~~.
\label{eq:t_SNT-jet}
\end{equation}

During the time at which the jet expands sideways, it appears to
a distant observer as if it were expanding into a medium of uniform 
density. This is due to the fact that 
during the time $r_\theta/c$ over which sideways expansion takes place, 
the jet radius does not increase significantly beyond $r_\theta$. Hence,
the density ahead of the shock can be approximated as being constant.

The temporal behavior of the afterglow flux at frequencies above
the synchrotron peak is
summarized for the different cases in Table~1, for a 
power-law energy distribution of the electrons with $p=2$.

\section{Discussion and Conclusions}

During the past year, there has been a growing consensus that at least some of
the observed GRBs involve collimated jets. At the same time, 
the observationally
inferred association of some GRBs with collapses of massive stars naturally
implies that in some cases the GRBs 
interact with an existing pre-outburst wind.

In the present work we have first examined the behavior of the afterglow in the
case of a collimated jet, 
including an interaction with a pre-existing wind. Our
results  can be summarized as follows:
\begin{enumerate}

\item We showed that while a break (followed by a steeper
decline) in the power-law decline is expected at $t\sim t_\theta$,
when the Lorentz factor decreases
below the inverse of the jet opening angle, a {\it flattening} in the light
curve is expected at $t\sim t_{\rm SNT}$,
when the flow approaches spherical symmetry. This new effect
is expected to occur about six months after the burst [for $z\ga1$; see
Eqs.~(\ref{eq:t_SNT}, \ref{eq:t_SNT-jet})].
The temporal behavior of afterglow flux at different stages of the jet
evolution is summarized in Table~1. 
We should also note that, contrary to a frequently quoted statement,
for a power-law energy distribution of the electrons with $p\approx2$ (as
observations indicate, e.g.\ Frontera et~al.\ 1999), the transition from a
relativistic to a non-relativistic expansion of a {\it spherical} fireball,
is {\it not} marked by a pronounced break in the decline (the transition is
only from 
$t^{-0.75}$ to $t^{-0.9}$ for frequencies where
 photons are emitted by
electrons cooling on time scales longer than the expansion time, while no 
transition is expected at a higher frequency; see Table~1).

\item While a measurement of the break time
$t_\theta$ alone does not allow for a direct determination
of the jet opening angle $\theta$, due to the dependence on the unknown
ratio of fireball energy to surrounding gas density [see Eqs.~(\ref{eq:theta},
\ref{eq:theta-jet})], a measurement of  {\it both\/} $t_\theta$ and the
flattening
time $t_{\rm SNT}$ allows to determine this ratio and therefore
allows for a direct determination of the jet opening angle $\theta$ 
[see Eqs.~(\ref{eq:t_SNT},\ref{eq:t_SNT-jet})].

\item The afterglow flux after flattening, i.e. once 
spherical symmetry is approached, is given
by Eqs.~(\ref{eq:nu_m}--\ref{eq:nu_c}), for the case of expansion into
a uniform density. Radio emission at 
the 1~mJy level, and optical emission at the $1\,\mu{\rm Jy}$ level
are expected at this stage.

\item On a timescale of days, the wind density is similar to typical ISM 
densities [see Eq.~(\ref{eq:n})], and therefore an interaction with a wind would 
give results that are not too different from the case of a uniform density.
In particular, steepening of the afterglow flux decline on a day time scale
implies a similar jet opening angle for the wind and ISM cases 
[see Eqs.~(\ref{eq:theta},\ref{eq:theta-jet})], a similar flattening time
[see Eqs.~(\ref{eq:t_SNT},\ref{eq:t_SNT-jet})], and hence similar fluxes
($\sim1$~mJy in the radio, $\sim1\,\mu{\rm Jy}$ in the optical)
at the onset of flattening.

\item The temporal decay indices for fast cooling electrons are similar 
in the wind interaction and uniform density cases (see Table~1). Consequently, 
differences may be detected by looking at slow cooling electrons (i.e.\ radio 
observations). However, the latter are affected by scintillation (e.g.\ Frail
et~al.\
1999).

\item Points (4) and (5) above imply that it would be difficult to
discriminate between the wind interaction and uniform cases through 
late-time observations. The two cases do give significantly different 
results though at $t\ll1$~day, when the densities are substantially 
different. Very early observations are therefore strongly favored.

\item Our results show that the case of an interaction with a wind is
{\it not\/} characterized by a significantly faster decline than the uniform
density case. Rather, to obtain a faster decline in the wind (non-collimated) 
case, one must assume a steeper electron index. Observations of a faster 
decline are therefore generally not indicative of an interaction with a wind.
\end{enumerate}

Examining the data on some GRBs in the light of the results presented above 
we note the following:
\begin{itemize}
\item[({\it i})] It does {\it not} appear that GRB~970228 and GRB~970508
involved 
highly collimated jets (since a steep, $t^{-2}$, 
decline has not been observed on a time scale of tens of days). 
Indeed, it is shown in Frail, Waxman \& Kulkarni (1999) that the
GRB~970508 radio data imply a wide-angle, $\theta\approx0.5$ jet expanding
into a uniform, $n\sim1{\rm\,cm}^{-3}$, density.\footnote{A model for 
the GRB~970508 afterglow, where a spherical fireball expands into a wind, has
been proposed by Chevalier \& Li (1999b). However,  Frail, Waxman \& 
Kulkarni (1999) have shown that this model is neither self-consistent 
nor consistent with the data.} Within the uncertainties it is difficult to
completely rule out an interaction with a wind for GRB~970228 (although 
that appears unlikely). Thus, while it is not altogether impossible that 
GRB~970228 was produced by a supernova (as suggested by Reichart, 1999, 
and Galama et~al.\ 1999, on the basis of the temporal properties and late 
spectral energy distribution), no sign of a supernova (similar to SN~1998bw) 
was found to underlie GRB~970508 (Fruchter 2000).

\item[({\it ii})] GRB~990123 and GRB~990510 may have had collimated jets. 
The relatively flat early decline in GRB~990510, however, argues against
interaction with a wind in this case, or suggests that the density in (at 
least some part of) the wind drops less steeply than $n\sim r^{-2}$. 
Furthermore, there is no significant evidence for a Type~Ic supernova 
underlying the GRBs (Fruchter et~al.\ 1999). It would be very interesting to 
search for the expected flattening in these two GRBs.
\item[({\it iii})] GRB~980519 and GRB~980326 may have involved collimated
jets. For this interpretation to be correct, however, the jet had to expand
sideways rather quickly. An underlying supernova has been tentatively 
identified in the case of GRB~980326 (Bloom et~al.\ 1999).
We should note that if the rapid sideways-expansion interpretation is correct,
then 
the jet should also get to the non-relativistic stage relatively early. Thus,
flattening 
of the light curve is expected in this case too.
\end{itemize}

Returning now to the question of the progenitors we note the following. It has
been argued, that the fact that in all cases in which an afterglow has been
unambiguously detected, the GRBs occurred {\it inside} galaxies, suggests
that these GRBs are not the result of NS--NS mergers (e.g.\ discussion in
Paczynski 1998; Livio et~al.\ 1998). This conclusion was based on the large
asymmetric kicks introduced in the latter systems by supernova explosions.
However, a detailed calculation showed that only $\sim15$\% of the GRBs are
expected to be found {\it outside} dwarf galaxy hosts (Bloom, Sigurdsson \&
Pols 1999). Thus, one might conclude that at least some of the observed GRBs
(with afterglows) could be the result of mergers of compact objects. The fact
that GRB~970228 and GRB~970508 showed neither a {\it clear} signature of a 
highly collimated jet, nor of an interaction with a wind,
could be taken as supporting evidence for this picture. As we explained
in \S2, NS--NS and NS--BH mergers are not expected to produce highly collimated
jets or interactions with a wind. Nevertheless, it remains true that NS--NS and
BH--NS mergers are generally expected to produce short duration bursts (and they
may provide the only model capable of producing extremely short [$<1$~sec]
bursts). In particular, even with the introduction of a reasonable disk
viscosity, it is difficult to see how the duration of the bursts could be
extended much beyond $\sim10$~sec (where most of the BeppoSAX sources are
found). Hence, we tentatively conclude that most of the long, relatively soft, 
GRBs with afterglows observed so far probably do not represent NS--NS (or 
BH--NS) mergers. Assuming this conclusion to be correct,  we may still be 
left with (at least) two main classes of progenitors for the (relatively)
long-duration, softer bursts: (1)~collapses of massive stars, and (2)~BH--helium
star mergers.  The mergers of BHs with white dwarfs, while capable (in 
principle) of producing long-duration bursts, appear to be too infrequent 
(e.g.\ Fryer, Woosley \& Hartmann 1999) to account for the observed bursts.

Neutron stars kicked by supernova explosions into their binary companion's 
envelope and transformed there into black holes (e.g.\ Bethe \& Brown 1998; 
but see Armitage \&  Livio 2000), have many similarities with collapses of 
massive stars and therefore will not be considered separately.

Our results suggest strongly that not all the long-duration GRBs originate 
in the collapses of massive stars, since there exist cases where no evidence 
was found for neither collimated jets not interactions with a wind. This 
prompts us to examine some of the aspects of models of the type of BH--helium 
star mergers.

In a BH--helium star merger, the helium core is dissipated to form a massive
disk (of radius $R_d\sim 
10^9$--$10^{10}$~cm, comparable to the core radius) around a spinning BH (e.g.\
Fryer \& Woosley 1998). Hence, there is no problem for this model (again, in
principle) to form collimated jets ($R_d/R_{\rm{CO}}\gg1$; see \S2). We
should note though, that in the same way that not all the Galactic BH X-ray
binaries produce highly collimated jets (see reviews of the properties by Chen,
Shrader \& Livio 1997, Mirabel \& Rodriguez 1998), it can (perhaps) be expected
that not all the GRBs will involve highly collimated jets.
Since in the common envelope phase which precedes the merger, matter may be
ejected preferentially in the orbital plane (e.g.\ Rasio \& Livio 1996;
Sandquist et~al.\ 1998), a significant interaction of the jet with the matter
ejected from the giant star's envelope may be avoided.

A comparison of the possible models with the discussion in points
({\it i})--({\it iii}) above therefore suggest the following (clearly at this 
stage very tentative) scenarios for progenitors: 
\begin{enumerate}
\item GRBs like GRB~980519 and GRB~980326 are produced by collapses of massive
stars, of the type that result in supernova explosions.
\item GRBs like GRB~990123 and GRB~990510 may be produced by BH--He star
mergers.
\item GRBs like GRB~9760228 and GRB~970508 may either be produced by BH--He
mergers which did not manage to collimate jets, or by a scenario presently
considered unlikely (like mergers of two compact objects).
\end{enumerate}

\acknowledgements
ML acknowledges support from NASA Grant NAG5-2678. EW acknowledges support 
from AEC Grant 38/99 and BSF Grant 9800343.

\newpage
\begin{table}
\caption{The temporal decay index $\alpha$ of afterglow flux, 
$f\propto t^{-\alpha}$, for an energy
distribution of the electrons $dn_e/d\gamma_e\propto\gamma_{e}^{-2}$
(values in brackets are for frequencies higher than the cooling frequency).}
\bigskip\bigskip
\small
\begin{tabular}{|c|cc|ccc|}
\hline
&\multicolumn{2}{c|}{Spherical Fireball} &\multicolumn{3}{c|}{Collimated Jet}\\
\hline
&relativistic &non-relativistic &before sideways &during sideways      &after
spherical\\
\noalign{\vspace{-0pt}}
&expansion &expansion        &expansion          &expansion (before   
&symmetry\\
\noalign{\vspace{-0pt}}
&           &                 &                   &spherical symmetry) &\\
\hline
uniform &&&&&\\
\noalign{\vspace{-0pt}}
density &$0.75\, [1]$ &$0.9\, [1]$ &$0.75\, [1]$ &$2\,[2]$ &$0.9\, [1]$\\
\hline
wind&&&&&\\
\noalign{\vspace{-0pt}}
($n\sim r^{-2}$) &$1.25\, [1]$ &$1.5\,[1]$ &$1.25\, [1]$ &$2\, [2]$ 
&$1.5\,[1]$\\
\hline
\end{tabular}
\end{table}

\end{document}